\documentclass{mem}
\usepackage{natbib}\usepackage{txfonts}\usepackage{balance}
\usepackage{graphicx}
\usepackage[a4paper]{hyperref}
\idline{00}{000}
\begin{document}

\title{Blazar microvariability at hard X-rays}


\author{L. \,Foschini\inst{1}, M. \,Gliozzi\inst{2}, E. \,Pian\inst{3}, G. \,Tagliaferri\inst{4}, F. \,Tavecchio\inst{4}, V. \,Bianchin\inst{1}, L. \,Maraschi\inst{4}, R. M. \,Sambruna\inst{5}, G. \,Di Cocco\inst{1}, G. \,Ghisellini\inst{4}, G. \,Malaguti\inst{1}, G. \,Tosti\inst{6} \and A. \,Treves\inst{7}}

\offprints{L. Foschini}
 
\institute{INAF/IASF-Bologna, Via Gobetti 101, 40129 Bologna, Italy\\
\email{foschini@iasfbo.inaf.it}
\and
George Mason University, Department of Physics and Astronomy, Mail Stop 3F3, 4400 University Drive, Fairfax, VA 22030, USA
\and
INAF/Osservatorio Astronomico di Trieste, Via G.B. Tiepolo 11, 34131 Trieste (Italy)
\and
INAF/Osservatorio Astronomico di Brera, Via Brera 28, 20121 Milano (Italy)
\and
NASA/Goddard Space Flight Center, Code 661, Greenbelt, MD 20771, USA
\and
Osservatorio Astronomico, Universit\`a di Perugia, Via B. Bonfigli, 06126 Perugia (Italy)
\and
Dipartimento di Scienze, Universit\`a dell'Insubria, Via Vallegio 11, 22100 Como (Italy)
}

\authorrunning{Foschini et al.}
\titlerunning{Blazar microvariability at hard X-rays}

\abstract{Blazars are known to display strong and erratic variability at almost all the wavelengths of electromagnetic spectrum. Presently, variability studies at high-energies (hard X-rays, gamma-rays) are hampered by low sensitivity of the instruments. Nevertheless, the latest generation of satellites (\emph{INTEGRAL}, \emph{Swift}) have given suggestions not yet fully confirmed of variability on intraday timescales. Some specific cases recently observed are presented and physical implications are discussed (e.g. NRAO $530$). The contribution that \emph{SIMBOL-X} could give in this research topic is also outlined.
\keywords{BL Lacertae objects: general -- quasars: general -- quasars: individual: NRAO $530$ -- X-rays: galaxies}
}
\maketitle{}

\section{Introduction}
According to the common paradigm, blazars are powered by a supermassive black hole with relativistic jets
extending from the centre to the outer space.  Relativistic effects play a dominant r\^{o}le in characterizing the observed emission, since the angle between the jet axis and the Earth's direction is small, generally less than $10^{\circ}$.

However, the gap between this generally accepted picture and a detailed physical model is still large.
An incomplete list of topics to be addressed in this research field, with particular reference
to the variability, should include: (\emph{i}) how jets are generated, collimated, and accelerated; (\emph{ii}) what is their composition (electrons, positrons, protons, ...); (\emph{iii}) what is their energy budget; (\emph{iv}) what is the origin of flares; (\emph{v}) what is the r\^{o}le of shocks; and many other unknowns related to the 
jet structure and dynamics.

To investigate these questions and -- from a more general point of view -- the nature of blazars, we need reliable sets of data covering energy ranges as broad as possible. Presently, this is not possible, particularly at high energies, because of technological limits. These problems become crucial as we probe short timescales, where the instruments sensitivity, rather than the time resolution, is the number one problem.

In the present essay, we want to address some of these problems commonly encountered when trying to investigate the shortest timescales with particular reference to the observations at hard X-rays. After this brief review, we will show how \emph{SIMBOL-X} can give an efficient answer to the present requirements in term of instrument capabilities.

\section{Variability at $\gamma-$ray energies}
In the MeV-GeV energy band, \emph{CGRO}/EGRET ($1991-2000$) changed our view of the $\gamma-$ray sky, particularly for blazars, but needed days to weeks integration time to get reliable measurement of flux of blazars in outburst. Only in a few exceptional cases it was possible to measure flux changes on hourly timescales, down to about 3 hours with PKS $1622-29$ (Mattox et al. 1997) or 8 hours for 3C $279$ (Wehrle et al. 1998). Great improvements are expected with GLAST\footnote{\texttt{http://www-glast.stanford.edu/}}, which will be launched at the end of January 2008, while AGILE\footnote{\texttt{http://agile.rm.iasf.cnr.it/}}, launched on April 2007, is still under the performance verification phase, so that its in-flight characteristics are still to be assessed.

At the highest energies (TeV), the new generation Cerenkov telescopes give nice results and in case of exceptional
events, like that of PKS $2155-304$ in July $2006$ (Aharonian et al. 2007) or Mkn $501$ in May--July $2005$ (Albert et al. 2007), have measured variability down to a few minutes . 

\section{Variability at X-ray energies}
In the energy band below $10$ keV, \emph{Chandra} and \emph{XMM-Newton} have opened new possibilities for X-ray astronomy. High sensitivity coupled with a large collecting area and long-period orbit ($48$ hours) make \emph{XMM-Newton} one of the best satellites for timing studies of extragalactic sources (e.g. S5 0716+71, Foschini et al. 2006a). In case of bright sources, it is possible to make variability studies with spectra resolved on timescales around $100$ s (e.g. Mkn 421, Ravasio et al. 2004).

Another important player in the theatre of operations is the X-ray telescope (XRT) onboard \emph{Swift}, which couples high sensitivity and time resolution together with a very fast repointing capability, making it possible to quickly follow the rise phase of outbursts or to perform long (several weeks) follow--up campaigns with short snapshots (e.g., PKS $2155-304$, Foschini et al. 2007). 

\section{Variability at hard X-ray energies}
In the hard X-rays, at energies above 10 keV, great advancements have been obtained by the \emph{INTEGRAL} instruments and by the Burst Alert Telescope (BAT) onboard \emph{Swift}. However, with the exception of a handful of very bright sources that are always within the reach of their sensitivity (e.g. 3C $273$, Bianchin et al. 2008), these instruments 
only detect blazars during outburst phases (e.g. S5 0716+71, Pian et al. 2005; 3C 454.3, Pian et al. 2006, Giommi et al. 2006). 

Variability studies are strongly limited by the instruments sensitivity and, despite the high fluxes reachable when the source is undergoing an active phase, it is possible to probe only daily timescales. However in the case of the exceptionally fast and bright flare of NRAO $530$ observed by \emph{INTEGRAL} (Foschini et al. 2006b), it was possible to measure flux changes over hourly timescales, but with no spectral information.

Slightly better results can be obtained with non-imaging instruments (\emph{RXTE}, \emph{Suzaku}/HXD), but there is the risk of source confusion.

\begin{figure*}[ht!]
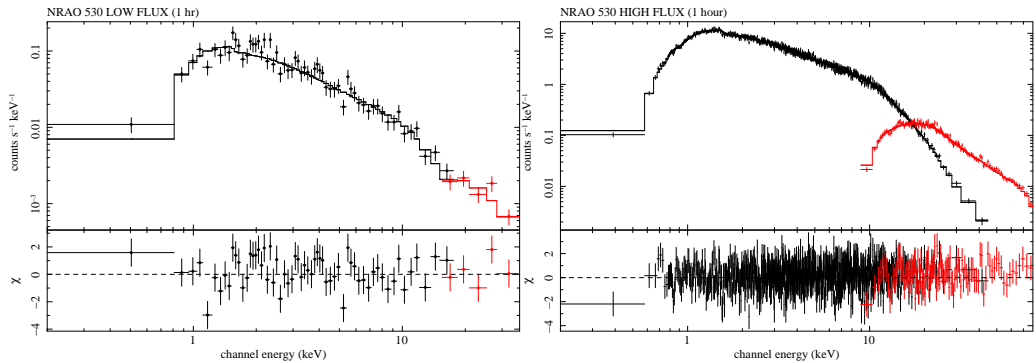

\resizebox{\hsize}{!}{\includegraphics[angle=270,clip=true]{foschini_f1a.ps}\includegraphics[angle=270,clip=true]{foschini_f1b.ps}}
\caption{\footnotesize Simulations of \emph{SIMBOL-X} observations of the blazar NRAO $530$ in quiescence (\emph{left}) and outburst (\emph{right}). The exposure was $1$ hour in both cases. See the text for details.
}
\label{sxsimulations}
\end{figure*}

\section{The case of NRAO $530$}
Since we will adopt NRAO $530$ as a case study, we briefly summarize the properties of its rapid variability event and we refer to Foschini et al. (2006b) for more details.

This event occurred on February $17$, $2004$, when \emph{INTEGRAL} was performing a deep exposure of the Galactic Centre. NRAO $530$ is usually below the sensitivity limit of the imager IBIS, but within about 1 hour the blazar reached a flux of about $2\times 10^{-10}$~erg~cm$^{-2}$~s$^{-1}$ in the $20-40$ keV energy band ($5\sigma$ detection in the fully-coded field of view) and then faded again below the detection limit. 

This flare was so short and intense to cast some doubts on its origin (it might have been produced by a
Galactic source within the $3'$ error circle of IBIS/ISGRI instead of the blazar). However, four observations with \emph{Swift}/XRT performed in $2006$ and $2007$ have shown that the only source present inside the IBIS error circle is NRAO $530$. Moreover, observations at other wavelengths supported the fact that this strong variability could be due to the blazar. Indeed, NRAO $530$ is known to display intense variability at almost all frequencies: e.g. optical (Webb et al. 1988), radio and $\gamma-$rays (Bower et al. 1997). Radio observations at $2$ cm with the VLBA (The MOJAVE Project, Lister \& Homan 2005) carried out a few days before the event (February $11$, $2004$), revealed a moderate increase of the polarization and after roughly 1 year radio maps showed a clear expansion of the radio jet. 

Possible explanations of this short hard X-ray flare (the first of this kind in a flat-spectrum radio-quasar type of blazar) are instabilities in the jet, which in turn might be due to one single shock, or an internal shock of two relativistic shells. In both cases, large Doppler factors would be necessary ($\delta \rightarrow 100$), while earlier radio observations suggest values around $\delta \sim 15$ (Bower \& Backer 1998). Recently, Homan et al. (2006) proposed a new method to measure the Doppler factor in blazar jets and found that $\gamma-$ray emitting AGN have the highest value of $\delta$. Specifically, NRAO $530$ has the highest value measured to date ($\delta \rightarrow 30-40$, MOJAVE Team, private communication), which makes it easier to explain the short hard X-ray flare.

\section{We need \emph{SIMBOL-X}}
The above described case is paradigmatic in the research field of short-term variability in blazars: it has high scientific impact, but the present technology strongly limits the available information. Particularly, no spectral slopes can be measured. 

The main problem in having a hard X-ray sensitivity suitable for this type of studies relies on the possibility of focussing X-rays at energies greater than $10$ keV. Presently, there are two ways to perform observations at $E> 10$ keV: collimators (\emph{RXTE}, \emph{Suzaku}/HXD) or coded-masks (\emph{INTEGRAL}, \emph{Swift}/BAT). The first method results in a sensitivity better than the second one, but it has no imaging capabilities. The coded-mask technique is presently the only method to obtain images at hard X-rays, but this means a sensitivity slightly worse than that available with collimators. 

The only way to have a meaningful improvement in hard X-ray sensitivity is to use focussing techniques. This is the main purpose of the mission \emph{SIMBOL-X} (Ferrando et al. 2008), expected to be launched in $2012-2013$. \emph{SIMBOL-X} is composed of two satellites (one with mirrors and the other with detectors) in formation flight and should operate in the $0.5-100$ keV energy band. The selected technology should allow an improvement of three orders of magnitudes in the $20-40$ keV energy band, making it literally possible to open a new observational window in a poorly explored band.

\section{Simulations}
In order to understand the potential of \emph{SIMBOL-X} for probing the hard X-ray microvariability, we have simulated the above mentioned case of the short intense flare of NRAO $530$. We have simulated 2 observations, one in quiescent state and one in flaring state, with an exposure time of 1 hour each.

For the quiescence, we adopted the spectrum measured by \emph{Swift}/XRT in the $0.3-10$ keV energy band (Kadler et al. 2007) and we extrapolate it to $E>10$ keV. This XRT spectrum is modelled with a power law with $\Gamma = 1.5\pm 0.3$ with absorption $N_{\rm H}=(5\pm 1)\times 10^{21}$~cm$^{-2}$. The $0.3-10$ keV flux was measured as $2.2\times 10^{-12}$~erg~cm$^{-2}$~s$^{-1}$. 

The outburst spectrum was modelled with the same parameters ($N_{\rm H}$, $\Gamma$), but with a greater normalization in order to have the extrapolated $20-40$ keV flux equal to the value measured by \emph{INTEGRAL} during the $2004$ flare.

The two simulated spectra are shown in Fig. \ref{sxsimulations}. The results of the simulations show that the absorption can be measured with $\approx 20\%$ accuracy in the quiescence and with $\approx 2\%$ in outburst. The errors on the photon index are $\Delta \Gamma = 0.1$ and $\Delta\Gamma =0.008$ for quiescence and outburst, respectively. 

This means that \emph{SIMBOL-X} will offer the unique possibility to follow the spectral evolution of this type of flares on hourly timescales.

\section{Conclusions}
The study of the high-energy short-term variability of blazars offers a great tool to understand the nature of these enigmatic sources. During the last $10$ years, several satellites were launched and improved significantly the available technology. However, the hard X-ray energy band is still one step behind with respect to the other energy bands, hampering the studies in this research field. 

The \emph{SIMBOL-X} mission will offer a valid and efficient answer to these problems. Simulations performed here show that with the foreseen improvement in sensitivity of three orders of magnitude in the $20-40$ keV energy band, it will be possible to follow the spectral and temporal evolution of short flares on hourly timescales.

Moreover, it will allow us to simultaneously study the soft and hard X-ray variability, sampling crucial part of the blazar spectral energy distribution, both for the synchrotron and the inverse Compton components.

\bibliographystyle{aa}

\end{document}